\documentclass{jltp}
\usepackage{graphicx}
\usepackage{latexsym}
\usepackage{amssymb}
\usepackage{psfrag}

\newcommand\RR{{\hbox{I\kern-.14em{R}}}}

\title{Transient growth of Kelvin waves on quantized vortices}

\author{C. F. Barenghi$^1$, M. Tsubota$^2$, A. Mitani$^2$
and T. Araki$^2$}

\address{$^1$School of Mathematics and Statistics, University of Newcastle,
Newcastle, NE1 7RU, UK\\
$^2$Physics Department, Faculty of Science, Osaka City University, Japan}

\runninghead{Barenghi, Tsubota, Mitani and Araki}{Transient growth of Kelvin
waves on
quantized vortices}

\begin{document}

\maketitle

\begin{abstract}
Quantized vortex lines in helium~II can be destabilized
by a sufficiently large normal fluid velocity which is parallel to the vortex
lines (Donnelly - Glaberson instability).  We study what
happens if the driving normal fluid is not steady but oscillates periodically
with time.  We show that in certain situations, although individual
perturbations decay for $t\to \infty$, a state of transient growth of
Kelvin waves is possible, which may explain observations of a small and
not reproducible longitudinal component of mutual friction.
\end{abstract}

\section{Introduction}

Recent work on the hydrodynamics of helium~II has highlighted
the importance of the Donnelly - Glaberson instability.
The instability takes the form of Kelvin waves
(rotating helical displacements of the vortex core) which grow
exponentially along quantized vortex lines.
If this happens, the length of the vortex lines increases
and energy is transferred from the normal fluid to the
superfluid\cite{Kivotides}.  The Donnelly - Glaberson instability 
was discovered\cite{Cheng,Glaberson} when a
small heat current was applied in the direction parallel to quantized
vortices in rotating helium~II. Figure 1 shows this rotating
counterflow configuration.
A channel is closed at one end and open to the helium bath at the
opposite end. At the closed end a resistor $R$ dissipates a known
heat flux $Q$. The heat is carried away from the resistor by the
normal fluid whilst the superfluid flows in the opposite direction
to assure that the net mass flux is zero. In this way a counterflow
velocity $V$ is set up which is proportional to the applied
heat flux.  If the channel is rotated
about its axis at angular velocity $\Omega$, in the absence of heat flux
there would be a uniform array of superfluid vortices of areal density
$L=2\Omega/\Gamma$ aligned along the rotation axis where $\Gamma$ is
the quantum of circulation. In the presence of a heat flux, the normal
fluid velocity ${\bf v}_n$
is parallel to the vortices and, if $Q$ is large enough,
the Donnelly - Glaberson instability takes place and Kelvin waves grow
exponentially with time. When the amplitude of the Kelvin waves becomes
of the order of the intervortex spacing $b\approx L^{-1/2}$,
vortex reconnections take place and,
according to a recent calculation \cite{Araki},
the uniform array of vortices may turn into a state of polarized turbulence.

It must be stressed that, although the rotating heat transfer
configuration of Figure 1 is the best to study the Donnelly -
Glaberson instability in a controlled way, there are other experiments
in which the Donnelly - Glaberson instability plays a key role.
For example, if helium~II is made turbulent using a towed grid\cite{Skrbek}
or rotating  propellers\cite{Tabeling}, both normal fluid and superfluid are
turbulent (the former consisting of eddies of various sizes
and strengths, the latter consisting of a disordered tangle of quantized
vortices).  In this case the instability may
take place throughout the turbulent flow, provided that
the component of ${\bf v}_n$ locally parallel to
a superfluid vortex filament exceeds a critical value \cite{ABC,Hulton}.

The aim of this paper is to remark that a similar Donnelly - Glaberson
instability can be excited if ${\bf v}_n$ is not steady
but varies harmonically with time. The natural way to do this is to modify
the configuration of Figure 1 and connect the resistor $R$ to an AC
(rather than a DC) power supply.  For the sake of brevity we shall
refer to the old and new configurations  as the
"DC instability" and the "AC instability" respectively.

The plan of the paper is the following: after deriving the DC instability
in Section 2, we describe two versions of the new instability in
Sections 3 and 4. The results are discussed in Section 5 in view of
previous experiments.

\section{The DC instability}

Let us consider a quantized vortex which points in the $z$ direction and
assume that its shape is perturbed in the form of a small amplitude
Kelvin wave of wavelength $\lambda$. A point on the vortex has Cartesian
components
\begin{equation}
{\bf s}(\xi)=[A\cos{(\phi)};A\sin{(\phi)};\xi]
\label{s}
\end{equation}
where $\phi=k\xi - \omega t$, $t$ is time, and $\omega$,
$A$ and $k=2\pi/\lambda$ are respectively the angular frequency, the amplitude
and the wavenumber of the wave.  Assuming $A k <<1$, it is easy to verify
that the parameter $\xi$ is the arc length. In fact
${\bf s}'=d{\bf s}/d\xi
=[-kA\sin(\phi);kA\cos(\phi);1]$ is the tangent to the curve at the point
${\bf s}$ and the length $\Lambda$ measured along the curve is
\begin{equation}
\Lambda=\int_0^{\xi} d\xi_0 \sqrt{
\vert {\bf s}'(\xi_0) \cdot {\bf s}'(\xi_0)\vert}
=\int_0^{\xi}d\xi_0\sqrt{k^2 A^2+1}\approx \xi.
\label{xi}
\end{equation}

In the absence of friction and using the local induction
approximation\cite{Saffman}, the self-induced velocity of the vortex at
the point ${\bf s}$ is
\begin{equation}
\frac{d{\bf s}}{dt}={\bf v}_{self}=\nu_s {\bf s}' \times{\bf s}'',
\label{lia}
\end{equation}
where $\nu_s=(\Gamma/(4\pi))\log{(b/a)}$ and
$a$ is the vortex core radius.  From (\ref{s}) and (\ref{lia}) we obtain
the dispersion relation of Kelvin waves
\begin{equation}
\omega \approx \nu_s k^2.
\label{omega}
\end{equation}

In the presence of friction the equation of motion is \cite{Schwarz}
\begin{equation}
\frac{d{\bf s}}{dt}={\bf v}_{self} +\alpha {\bf s}' \times
({\bf v}_n - {\bf v}_s - {\bf v}_{self})
\label{dsdt}
\end{equation}
where $\alpha$ is a known mutual friction coefficient and ${\bf v}_n$ and
${\bf v}_s$ are imposed normal fluid and superfluid  velocity fields.
Choosing the
$z$ axis along the counterflow channel as in Figure 1,
we have ${\bf v}_n-{\bf v}_s=[0;0;V]$ where $V$ is the applied
counterflow velocity. Allowing the amplitude of the wave to change with
time and
using~(\ref{omega}) we obtain
\begin{equation}
\frac{dA}{dt}=\alpha (k V -\nu_s k^2 )A,
\label{dadt:0}
\end{equation}

If $V$ is constant the solution of equation~(\ref{dadt:0}) is
\begin{equation}
A(t)=A(0)\exp{(\sigma t)},
\label{a:0}
\end{equation}
where $A(0)$ is the initial infinitesimal amplitude of the wave due to
thermal or mechanical vibrations and
$\sigma=\alpha (kV-\nu_s k^2)$ is the growth rate.
The instability occurs if $\sigma>0$. Therefore a Kelvin wave of
wavenumber $k$ is unstable in the region $0<k<k_0$ where $k_0=V/\nu_s$
and the maximum growth rate $\sigma_c=\alpha V^2/(4\nu_s)$  occurs at
wavenumber $k_c=V/(2\nu_s)=k_0/2$. The temperature of the helium bath
determines $\alpha$ and the applied heat current determines $V$, so
the Kelvin wave which grows exponentially faster and dominates all the other
waves has wavelength $\lambda_c=4\pi\nu_s/V$.

\section{The AC instability: second sound}

In second sound the quantity $V$ reverses sign during a cycle.
This can be achieved by using Nuclepore
transducers, which are oscillating membranes covered by holes which
are so small (sub-micron size) that the viscous normal fluid cannot move
through them. In this way an oscillating counterflow
velocity is set up and  Eq.~(\ref{dadt:0}) becomes
\begin{equation}
\frac{dA}{dt}=\alpha(kV_0 \cos(\omega_0 t) - \nu_s k^2)A,
\label{dadt:1}
\end{equation}
Letting $t'=\omega_0 t$ and $\mu=\nu_s k/V_0$ the solution of
Eq.~(\ref{dadt:1}) is
\begin{equation}
A(t')=A(0) e^{f(t')},
\label{a:1}
\end{equation}
where
\begin{equation}
f(t')=\frac{\alpha k V_0}{\omega_0} [\sin(t') -\mu t'],
\label{f}
\end{equation}

The amplitude $A$ of the perturbation grows if $f>0$, but, since
$\vert \sin(t')\vert \le 1$, we have $f<0$  for sufficiently long times.
We conclude that infinitesimal perturbations of any wavenumber $k$
decay exponentially for $t\to \infty$. However,
although there are no instabilities for $t\to \infty$,
for a limited initial time helical waves can actually increase their
amplitude. This "transient growth" takes place in the time interval
such that $\sin(t') >\mu t'$ that is to say for
\begin{equation}
t<\frac{V_0}{\nu_s k \omega_0}
\end{equation}
The effect can be significant (it can last for an
appreciable time)
at low second sound frequency ($\omega_0<<1$),
high second sound amplitude ($V_0>>1$) and long wavelength ($k<<1$),
because $\nu_s \approx 10^{-3}\rm cm^2/sec<<1$
(taking $a_0\approx 10^{-8}\rm cm$
and $\Omega \approx 1\rm rad/sec$). Note that the longest wavelength
depends on the size of the channel.
The important point to notice is that a great range of
wavenumbers can be affected by transient growth.
Since mechanical and thermal perturbations are created all the time, if
these perturbations grow before decaying to zero at $t\to \infty$,
then each vortex line will have a certain amount of
small amplitude Kelvin waves, depending on the particular set up of the
experiment ($\omega_0$, $V_0$ and the size of the channel)
By attenuating the second sound which
propagates in the direction along the vortices, these Kelvin waves
may lead to think that there exists a longitudinal coefficient of
mutual friction.

\section{The AC instability: oscillating counterflow}

Perhaps the simplest way to produce a time dependent
heat flux $Q$ is to apply an alternating voltage to the resistor $R$
which generates the heat current. Since $Q$ is proportional to the square of
the applied voltage, Eq.~(\ref{dadt:1}) must be replaced by
\begin{equation}
\frac{dA}{dt}=\alpha [k V_0 \cos^2(\omega_0 t) -\nu_s k^2]A,
\label{dadt:3}
\end{equation}
Preceding as before we have
\begin{equation}
\frac{dA}{dt'}=\frac{\alpha k V_0}{\omega_0}[\cos^2(t')-\mu]A,
\label{dadt:3b}
\end{equation}
and the solution is
\begin{equation}
A(t')=A(0)e^{f(t')},
\label{a:3}
\end{equation}
\noindent
where
\begin{equation}
f(t')=\frac{\alpha k V_0}{2 \omega_0}[(1-2\mu)t' + \frac{1}{2} \sin(2t')],
\label{f:2}
\end{equation}
\noindent
The perturbations grow if $f(t')>0$. The long term behaviour of $f(t')$
depends on the first term in Eq.~(\ref{f:2}) which increases with $t'$,
not on the second which is always less than unity. If the first term is
positive ($1-2 \mu > 0$) then $A$ grows for $t \to \infty$;
this happens for $\mu<1/2$, that is for
$0<k<\frac{V_0}{2\nu_s}$
which is the same condition as for the DC instability with $V_0$
replaced by $V_0/2$. Transient growth is possible
for a limited time even if $1-2\mu<0$ provided
that is for $t<1/(2\omega_0\vert1-2\mu\vert)$.
\bigskip

\section{Conclusion}

We have shown that there is an "AC" version of the "DC" Donnelly - Glaberson
instability, and that can it be realized in a controlled way
in the laboratory, either using second sound Nuclepore transducers
or an oscillating counterflow. We have determined the
boundaries between the stable and the unstable regions and found
that in the linearly stable regions, although each individual perturbation
decays as $t\to \infty$, a state of transient growth is possible
which depends on the frequency and the
amplitude of the second sound wave and on the length of the channel.
This state of transient growth of Kelvin waves would produce attenuation of
second sound in the direction parallel to the quantised vortices and
would be interpreted as evidence for a longitudinal component of mutual
friction.  The possibility of the existence of a longitudinal friction
coefficient (called $B''$, $B_z$ or $\gamma$) was
considered in the earlier helium~II literature\cite{Khalatnikov}.
Experiments\cite{Snyder} concluded that $B''$, if it exists
at all, is very small (at the most $500$ times smaller than $B$),
is not reproducible and may depend on the amplitude of the second sound
wave used to measure it, which is consistent with our finding.
We plan to continue this investigation using numerical simulations.

C.F.B. is grateful to the Royal Society for supporting this study.

\begin{figure}
\begin{center}
\includegraphics[angle=0,width=0.2\textwidth]{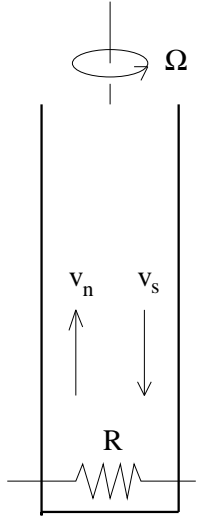}
\caption{Rotating counterflow channel configuration.}
\label{fig1}
\end{center}
\end{figure}

\end{document}